\documentclass[structabstract]{aa} 

\usepackage{graphicx}
\usepackage{amsmath}
\usepackage{txfonts}
\usepackage{natbib}
\usepackage{xspace}
\usepackage{textcomp}
\usepackage[acronym,nowarn]{glossaries}
\usepackage[textsize=tiny,textwidth=5em]{todonotes}
\usepackage{siunitx}
\makeatletter
\newlength{\figwidth}
\if@referee
\newcommand{\hl}[1]{#1}
\else
\newcommand{\hl}[1]{#1}
\fi
\makeatother

\newcommand{\be}{\begin{equation}}
\newcommand{\ee}{\end{equation}}
\newcommand{\bs}{\begin{subequations}}
\newcommand{\es}{\end{subequations}}

\newacronym{MHD}{MHD}{magnetohydrodynamic}
\newacronym{SNR}{SNR}{supernova remnant}
\newacronym{ISM}{ISM}{interstellar medium}

\newcommand{\alfven}{Alfv\'{e}n\xspace}
\newcommand{\alfvenic}{Alfv\'{e}nic\xspace}
\newcommand{\padian}{\textsc{Padian}\xspace}

\DeclareSIUnit\kms{\kilo\meter\per\second}
\DeclareSIUnit\parsec{pc}
\DeclareSIUnit\lightyear{ly}
\DeclareSIUnit\year{y}
\DeclareSIUnit\century{century}
\DeclareSIUnit\erg{erg}

\newcommand{\diff}{\mathrm{d}}

\renewcommand{\c}{\mathrm{c}}

\begin{document}
\title{Cosmic-ray acceleration at collisionless astrophysical shocks using Monte-Carlo simulations}
\author{M. Wolff \and R.\,C. Tautz}
\institute{Zentrum f\"ur Astronomie und Astrophysik, Technische Universit\"at Berlin, Hardenbergstra\ss{}e 36, D-10623 Berlin, Germany\\\email{robert.c.tautz@gmail.com}}

\date{Received \today; accepted April 4, 2063}

\abstract
{The diffusive shock acceleration mechanism has been widely accepted as the acceleration mechanism for galactic cosmic rays. While self-consistent hybrid simulations have shown how power-law spectra are produced, detailed information on the interplay of diffusive particle motion and the turbulent electromagnetic fields responsible for repeated shock crossings are still elusive.}
{The framework of test-particle theory is applied to investigate the effect of diffusive shock acceleration by inspecting the obtained cosmic-ray energy spectra. The resulting energy spectra can be obtained this way from the particle motion and, depending on the prescribed turbulence model, the influence of stochastic acceleration through plasma waves can be studied.}
{A numerical Monte-Carlo simulation code is extended to include collisionless shock waves. This allows one to trace the trajectories of test particle while they are being accelerated. In addition, the diffusion coefficients can be obtained directly from the particle motion, which allows for a detailed understanding of the acceleration process.}
{The classic result of an energy spectrum with $E^{-2}$ is only reproduced for parallel shocks, while, for all other cases, the energy spectral index is reduced depending on the shock obliqueness. Qualitatively, this can be explained in terms of the diffusion coefficients in the directions that are parallel and perpendicular to the shock front.}
{}

\authorrunning{Wolff \& Tautz}
\titlerunning{Cosmic-ray acceleration at astrophysical shocks}
\maketitle

\section{Introduction}

For many decades now, astrophysicists have investigated the origin of cosmic rays, the high-energy charged particle radiation that permeates space. Various experiments, both earthbound and space-borne, measure a characteristic cosmic-ray energy spectrum, which follows a (more-or-less) universal power law. The individual particles have kinetic energies above several \si{\mega\electronvolt}, and even energies up to more than \SI{E20}{\electronvolt} have been detected experimentally, as shown by air-shower experiments \citep[e.\,g.,][]{abraham2008upper,let11:uhe}. Overall, the data come from a multitude of astrophysical experiments that have been conducted over past decades. The understanding of the origin of the distinct shape of the spectrum continues to be an active field of study, and new experiments like the Square Kilometer Array \citep[SKA, e.\,g.,][]{fal04:uhe} are being assembled.
%

While ineffective on its own, the original \citet{fermi1949origin} acceleration mechanism inspired later research, leading to the development of the so-called diffusive shock acceleration mechanism or of first-order Fermi acceleration \citep{axford197715th,krymsky1977,blandford1978particle,bell1978acceleration}. This theory outlines an effective acceleration mechanism for charged particles at magneto-hydrodynamic shock waves. Such shock waves of varying sizes are ubiquitous throughout space, from solar winds through supernova remnants (SNR) to relativistic shocks in exotic cosmic objects, such as active galactic nuclei or pulsars (see \citealt{bal13:sho} for an introduction).

Thanks to the efficiency of the diffusive shock acceleration mechanism and the existence of plausible acceleration sites, it became accepted as the de-facto standard acceleration mechanism \citep[cf.][]{abd10:acc,hel12:snr}. There are, however, still many open questions that cannot be answered entirely yet.
\hl{These include direct proof for acceleration at the shock, the injection problem, and the origin of ultra-high energy cosmic rays has not yet been understood \citep[e.\,g.,][]{abb14:ani}.}
%
%
In general, the complexity of the physical problem at hand requires the application of multiple approximations to find results both analytically and numerically.
%

The work presented here will concentrate on non-relativistic shock waves in supernova remnants \citep[for a review see, e.\,g.,][]{rey08:snr,bla13:gcr}. Based on the numerical model used in the \padian code \citep{tautz2010new}, a test-particle simulation is extended to include shock waves. This approach is then applied in order to investigate the effect of diffusive shock acceleration by inspecting the obtained cosmic-ray energy spectra. This relatively simple test-particle model is compared to theory, on one hand, and the results of \citet{caprioli2014simulations}, on the other hand, who applied a sophisticated kinetic hybrid model to investigate particle acceleration at non-relativistic astrophysical shocks. Finally, the work presented here looks at the influence of the model of turbulence on the acceleration process. Most notably, a simple magnetostatic model is compared to an extended model based on \alfven waves, following the research of \citet{schlickeiser2002cosmic} on the transmission of these waves through a parallel shock front.

This article is organized as follows. In Sec.~\ref{acc}, a brief outline of diffusive shock acceleration is given, followed by a summary of the numerical model in Sec.~\ref{model}. The results are presented in Sec.~\ref{results}. Section~\ref{summ} provides a brief summary and a discussion of the results.

\section{Diffusive shock acceleration}\label{acc}

Shock waves are a special case of a general physical phenomenon, where the physical properties of a (magneto-)hydro\-dynamic environment change discontinuously across a surface area. They are characterized by being propagating structures; i.\,e., they have a non-zero mass flux across the discontinuity surface, which is commonly referred to as the shock front. In reality, a shock front is an irregular surface area and the discontinuity will occur over a certain small, but non-zero, distance. Any realistic description of such a system would be highly complex. Instead, one often approximates the shock front to be one-dimensional, planar, and to have infinitesimally small thickness. This simplification is justified on any scale far smaller than the total scale of the shock wave but much larger than the shock front's thickness.
%

Shock waves with very high upstream Mach numbers $M_1 \gg 1$ are called \textit{strong shock waves}. In this limit, the compression ratio $r$, i.\,e. the ratio between the upstream density $\rho_1$ and the downstream density $\rho_2$, can be written as \citep[e.\,g.,][Sec.~16.2.2.1]{schlickeiser2002cosmic}
\begin{equation}\label{eq:compression-ratio}
r = \frac{\rho_2}{\rho_1} = \frac{v_1}{v_2} = \frac{\gamma + 1}{\gamma - 1 + 2 \beta /M_1^2} \approx \frac{\gamma + 1}{\gamma - 1}.
\end{equation}
Here, the plasma $\beta$ is the ratio of the plasma pressure to the magnetic pressure, which will be assumed to be negligible small compared to the square of the shock Mach number $M_1^2$. The right-hand side is written in terms of the ratio of specific heats of the gas, $\gamma$. The compression ratio reaches its minimum of $r = 4$ for a monoatomic gas.
%
%
An example for a strong shock wave is the shell of a \gls{SNR}, which expands with a velocity in the order of \SI{E4}{\kms}. Compared to the speed of sound, the strong shock approximation is plausible.

Shocks with a velocity $V_\text{shock}$ larger than the \alfven speed $V_\text{A}$, which can be approximated to lie in the range of \SIrange{30}{50}{\kms} in the interstellar medium \citep[cf.][]{kan13:snr}, are said to be \textit{super-\alfvenic}. These \textit{fast shocks} are common elements of astrophysical environments. They can be found for example in \gls{SNR}s, the relativistic bulk outflow from active nuclei, pulsar wind nebulae, or gamma-ray bursts.
%

Averaging over particles crossing the shock at arbitrary angles and taking into account that, due to scattering processes both in the upstream and the downstream regions, the velocity vector will be quickly randomized, a particle's energy gain for a complete round-trip is given by
\begin{equation}
\left \langle \frac{\varDelta E}{E} \right \rangle= \frac{4}{3} \frac{V_\text{shock}}{\c}.
\end{equation}
\citet{bell1978acceleration} showed that only downstream are a small fraction of particles advected from the acceleration region of a non-relativistic strong shock. This leads to a probability for the particle to remain within the acceleration region of $P = 1 - V_\text{shock}/\c$. The resulting energy spectrum power-law is then given by
\begin{equation}
\diff N(E) \propto E^{-2} \diff E.
\end{equation}

Looking at \gls{SNR} shells as potential sites for diffusive shock acceleration, \citet{blandford1978particle} estimate that at least $10\%$ or even up to $50\%$ of the energy of a \gls{SNR} in the typical standard Sedov solution with $E = 10^{51}\,$ergs can be used for particle acceleration. The limiting factor is rather the maximum size of the \gls{SNR} shock front, which it reaches at the end of the so-called \textit{adiabatic Sedov-Taylor expansion}.
%

At the upper end of the cosmic-ray spectrum, particles are measured with such extreme energies that their gyroradii would exceed the dimensions of a \gls{SNR}. Additionally, the time spent by particles in the vicinity of the shock front plays a role: The faster a particle becomes, the faster it will be ejected from the acceleration zone. \citet{hillas1984} researched this topic extensively in the context of ultra-high-energy cosmic rays. By equating the particle's gyroradius with the dimension $L$ of an acceleration site, he finds a plausible upper bound to the maximum energy a particle can be accelerated to as
\begin{equation}\label{eq:hillas-criterion}
E_\text{max} = Z \text{e} B V_\text{S} L.
\end{equation}
This purely geometrical consideration is nowadays called the \textit{Hillas criterion} \citeyearpar{hil84:hil}, and applies to any kind of acceleration site. For a single nucleon with $Z = 1$ and a supernova with a size $L = \SI{100}{\parsec}$ and velocity $V_\text{S} = \SI{E4}{\kms}$ in a magnetic field of strength $B = \SI{E-10}{\tesla}$, the above equation yields an upper limit to the particle energy of around \SI{3E15}{\electronvolt}. Cosmic rays consisting of heavier atoms with more nucleons, such as iron, could potentially be accelerated to even higher energies of around \SI{E18}{\electronvolt} in the strong magnetic fields present in young \gls{SNR} \citep[sec. 17.5.3]{longair2011high}. Beyond that regime, either different acceleration mechanisms must play a role, or other cosmic objects become the relevant acceleration sites. The Hillas plot \citep[cf.][]{vuk07:hil,sha09:hil} shows a straight line for the solution of the Hillas criterion for a proton with $E_\text{max} = \SI{E20}{\electronvolt}$. Of the various potential accelerations sites shown in the plot, only those above the line could possibly accomplish such an acceleration.

Another open question is the origin of the initial particles. Both versions of the Fermi acceleration shown above assume that the induced particles already have relativistic velocities. Particles without a sufficiently high initial energy would not be accelerated due to ionization losses. This problem is called the \textit{injection problem}. Various theories have been proposed to overcome this issue, which have been summarized in the review of \citet{malkov2001modern}. Of particular interest to the work presented here is the ``thermal leakage'' scenario, where the generation of \alfven waves by accelerated particles is proposed. These waves could on one hand self-regulate the thermal particle injection and on the other hand lead to efficient pitch-angle scattering, confining accelerating particles to the shock wave and thereby increasing the efficiency of the diffusive acceleration mechanism \citep{malkov1998ion,winske1988magnetic}. Based on this idea, \citet{caprioli2014simulations} used a Maxwell-
Boltzmann distributed injection profile and could reproduce an efficient acceleration process, leading to a power spectrum with a spectral index of $-1.5$ in conformance with the theory under the assumption of non-relativistic particle speeds.

\section{Numerical model}\label{model}

In the \padian code, a Monte-Carlo method is applied to calculate the transport parameters of cosmic-ray test-particles. The particle equation of motion---the Newton-Lorentz equation---is solved numerically over a defined period of time, e.\,g., several thousand gyro periods.
%
%
This procedure is repeated for a sufficiently large number of particles, until,
%
%
the individual results are combined and statistically evaluated to reach an overall result.




\subsection{Magnetic field}\label{sec:method-magnetic-field}

\padian splits the magnetic field into two components: the turbulent component $\delta \vec{B}$ and the mean magnetic field $\vec{B}_0$, which models the large-scale magnetic background field. In the work presented here, a uniform constant magnetic background field directed along the $z$-axis is applied.

The turbulent component $\delta \vec{B}$ is calculated for every spatial coordinate $(x,y,z)$ and is not discretized. A detailed description of the model of magnetic turbulence applied in \padian is given in \citet{tautz2010new} and \citet{tautz2013numerical}. The following will reiterate the most important aspects of it.

Based on the work of \citet{shalchi2009random}, the power spectrum of the turbulence is modeled by a modified kappa distribution
\begin{equation}\label{eq:model-turbulence-energy-spectrum}
G(k) = \frac{(l_0 k)^q}{\left [ 1 + (l_0 k)^2 \right]^{(s+q)/2}},
\end{equation}
where $q$ and $s$ define the energy range and inertial range spectral indices, respectively. The transition between these two ranges occurs at the so-called \textit{bend-over scale} $l_0$, which will be used for normalization purposes as discussed below. For $q = 0$ and $s = 5/3$ one obtains the  Kolmogorov spectrum of turbulence with a flat energy range. An important alternative to the above spectrum has been given by \citet{iroshnikov1964turbulence} and \citet{kraichnan2003small}, who extended the original theory to fluids carrying a magnetic field: In their model, which is nowadays called Iroshnikov-Kraichnan turbulence, they assume locality of interactions, weak interactions, and isotropy. For turbulence in a plasma influenced by a uniform magnetic field they then arrive at an energy spectrum of the form
\begin{equation}
E(k) \propto (\epsilon V_\text{A})^{1/2} k^{-3/2},
\end{equation}
where $V_\text{A}=B_0/\sqrt{4\pi\rho}$ denotes the \alfven velocity. Numerous experiments successfully reproduced either form of the energy spectrum when investigating turbulent systems \citep{lrsp-2013-2}. But both of these theories, and many others, assume an isotropic turbulence which is contradicted by many observational results, in particular in the solar wind. More recently, \citet{goldreich1997magnetohydrodynamic} presented another model that takes the anisotropy in the turbulence of astrophysical \gls{MHD} plasmas into account. However, no model has been found which can explain all experimental data nor predict the results of numerical simulations. A recent review on this matter is given by \citet{schekochihin2007turbulence}.

\padian furthermore follows the ideas of \citet{giacalone1999transport} and generates the turbulent magnetic field by superposing multiple plane waves with random phase angles and directions of propagation. According to \citet{batchelor1953theory}, one can model isotropic, spatially homogeneous turbulence this way, when using a large number of waves modes. For an individual mode, the magnetic field can then be calculated with \citep{tautz2013numerical}
\begin{equation}
\delta \vec{B}_n = \hat{\zeta}_\perp A(k_n)\,\cos\left(k_n \zeta + \psi_n\right),
\end{equation}
where $A(k_n) \propto \sqrt{G(k_n)\,\varDelta k_n}$ is the amplitude function, which is defined by the energy spectrum (cf. Eq.~\eqref{eq:model-turbulence-energy-spectrum}). The unit vector $\hat{\zeta}_\perp$ is perpendicular to the direction of $\zeta$, which in turn is obtained by applying a random rotation matrix to the spatial coordinate $(x, y, z)$ for each wave mode $k_n$. Additionally, each wave mode has a random phase given by $\psi_n$.

The real part of a summation over a number of plane wave modes $N$ then yields the total turbulent magnetic field as
\begin{equation}\label{eq:model-magnetic-field}
\delta \vec{B} = \Re \sum\limits^N_{n=1} \delta \vec{B}_n.
\end{equation}

\subsubsection{\alfven{}ic turbulence}\label{sec:model-alfven}

The turbulence model introduced above can be extended to support propagating \alfven waves \citep[e.\,g.,][]{mic96:alf,tautz2010new}. This is achieved by modifying the turbulent magnetic field components of individual plane wave modes $\delta \vec{B}_n$ in Eq.~\eqref{eq:model-magnetic-field} with an oscillation frequency, yielding
\begin{equation}\label{eq:model-magnetic-field-alfven}
\delta \vec{B}_n = \hat{\zeta}_\perp A(k_n)\,\cos\left[k_n \zeta -\omega(k_n)t + \psi_n\right].
\end{equation}
Here, $\omega(k_n)$ denotes the dispersion relation of the \gls{MHD} wave, which, for \alfven waves, is given by $\omega(k_n) = \pm V_\text{A} k_\parallel$. It has to be noted that this model is only valid in absence of any shock effects, i.\,e. in the upstream region. Sec.~\ref{sec:model-alfven-shock} below discusses the downstream region.

Optionally, \padian can also include the effect of the electric field components of the \alfven waves. These are perpendicular to the magnetic field components and obtained from applying the Faraday induction equation
\begin{equation}
\vec{B} = \frac{\text{c}}{\omega} \vec{k} \times \vec{E}
\end{equation}
by taking the polarization properties of \alfven waves into account.

\subsection{Shock wave}

In order to investigate the effects of diffusive shock acceleration, \padian was extended to model a one-dimensional \gls{MHD} shock front: The ambient plasma will flow through the shock rest frame with a defined velocity toward the shock front. At the position of the shock front, a discontinuity in the magnetic field and ambient gas velocity is added according to the Rankine-Hugoniot jump conditions (see, e.\,g., \citealt{schlickeiser2002cosmic}, Chpt.~16 and \citealt{longair2011high}, Chpt.~11). Note that the back reaction of the cosmic-ray particles on the shock wave will be neglected \citep[cf.][]{riq10:amp}.

\padian internally works with dimensionless parameters, which simplifies the support for relativistic particles and improves the generality of the results. The time $t$ is normalized with the relativistic gyrofrequency to $\tau=\varOmega_\text{rel}t=qB_0t/(\gamma mc)$ with $\gamma$ the relativistic Lorentz factor. In place of the momentum, \padian instead uses the particle's rigidity $R$ normalized to the turbulence bend-over scale $l_0$, as given by $R=R_\text{L}/\ell_0=v/(\varOmega_\text{rel}\ell_0$.

Using these parameters, the equation of motion is solved in the gas rest frame instead of in the observer frame. According to Faraday's law, the flow of the ambient plasma will also induce an electric field
\begin{equation}\label{eq:efield_induct}
\vec{E}_\text{ind.} = - \frac{1}{\text{c}} \vec{R}_\text{flow} \times \vec{B},
\end{equation}
where $\vec{R}_\text{flow}$ represents the flow rigidity, and $\vec{B}$ the total magnetic field. In the shock frame of reference, which is used in the simulations presented in Sec.~\ref{results}, this will influence the particle rigidity even without them crossing the shock front. The effect of the flow of the ambient plasma gas on the test particles is taken into account by adapting the Newton-Lorentz equation to include the inductive electric field from Eq.~\eqref{eq:efield_induct}, yielding
\begin{equation}
\frac{\diff}{\diff \tau} \vec{R} = \left (\vec{R} - \vec{R}_\text{flow} \right) \times \vec{B}.
\end{equation}
The relative rigidity $\vec{R}_\text{flow}$ follows from the Lorentz transformation into the gas rest frame. It is always parallel to the shock normal, and scales analogously to the gas velocity across the shock front. For a given value of the shocks rigidity $\vec{R}_\text{S}$ and compression ratio $r$, one can calculate the relative rigidity upstream and downstream of the shock front using
\bs
\begin{align}
\vec{R}_\text{rel,1} &= \vec{R}_\text{S} , \\
\vec{R}_\text{rel,2} &= \frac{1}{r} \vec{R}_\text{S}.
\end{align}
\es
Upstream, the magnetic field $\vec{B}_1$ is calculated directly according to Eq.~\eqref{eq:model-magnetic-field}. Downstream, the tangential magnetic field component $B_\text{2,t}$ is scaled with the compression ratio as defined by the Rankine-Hugoniot equations, while the normal component $B_\text{n}$ is continuous, leading to
\bs
\begin{align}
B_\text{2,n} &= B_\text{n}\\
B_\text{2,t} &= r B_\text{t}. \label{eq:model-magnetostatic-downstream}
\end{align}
\es
This simple model allows the investigation of parallel, oblique, and perpendicular fast, strong shock waves.

\subsubsection{Interaction with \alfven Waves}\label{sec:model-alfven-shock}

\alfven waves interact with parallel shock fronts, which process was described in detail by \citet{campeanu1992,schlickeiser1993,vainio1998,vainio1999}, and \citet{vainio2003alfven}. A review of this combined work is given in \citet[Chpt.~16.3]{schlickeiser2002cosmic}.

The important result for the work presented here are the transmission coefficients at constant wavenumber $k$. They are found to be \citep[eq. 16.3.26f]{schlickeiser2002cosmic}
\bs
\begin{align}
T_k &= \frac{\sqrt{r} + 1}{2 \sqrt{r}} \left( r \frac{M + H_{\text{c},1}}{M + \sqrt{r} H_{\text{c},1}} \right)^{(s+1)/2} \\
R_k &= \frac{\sqrt{r} - 1}{2 \sqrt{r}} \left( r \frac{M + H_{\text{c},1}}{M - \sqrt{r} H_{\text{c},1}} \right)^{(s+1)/2} .
\end{align}
\es
Here, $r$ is the shock compression ratio and $s$ is the inertial range spectral index with values $1 < s < 2$. For the work presented here, it is set to $5/3$ (cf. Sec.~\ref{sec:method-magnetic-field}). Finally, the equation above includes the normalized cross helicity state $H_{\text{c}}$ of the waves, which is given by
\begin{equation}
H_\text{c} = \frac{(\delta B_\text{f})^2 - (\delta B_\text{b})^2}{(\delta B_\text{f})^2 + (\delta B_\text{b})^2}.
\end{equation}
In \padian, only forward moving waves are injected upstream so that $H_{\text{c},1} = +1$. There, the turbulent magnetic field is calculated using Eq.~\eqref{eq:model-magnetic-field-alfven}. Downstream of the shock front, the formula is adapted to take both the forward and backward traveling \alfven waves with the corresponding transmission coefficient into account, leading to
\begin{align}
\delta \vec{B}_2 &= \Re \sum\limits^{N_\text{m}}_{n=1} \delta \vec{B}_n\left [ T_k \sin \left( k_n \zeta - \omega(k_n) t - \varPsi_n \right )\right.\nonumber\\
&+ \left.  R_k \sin \left( k_n \zeta + \omega(k_n) t - \varPsi_n \right ) \right ].\label{eq:model-magnetic-field-alfven-downstream}
\end{align}
This model for the downstream turbulent magnetic field already leads to a significant increase in the magnetic field strength, compared to the upstream values. As such, contrary to the magnetostatic model of turbulence of Eq.~\eqref{eq:model-magnetic-field}, no artificial compression following Eq.~\eqref{eq:model-magnetostatic-downstream} must be applied.

Finally, it is important to state that this model is currently limited to parallel shock configurations. \citet{vainio1999} explicitly mention in their work that it might not be easily extensible to oblique shock configurations: ``Note that our treatment holds only for unidirectional circularly polarized upstream wave fields, since variations in the magnitude of the perpendicular magnetic field component may induce motion of the shock front itself from its equilibrium position \citep[e.\,g.,][]{achterberg1986}. The situation gets even more complicated, if the upstream wave vectors are not aligned with the shock normal, since the assumption of a planar shock is then probably violated, also. This is why we do not expect our model to be generalized to oblique shocks very easily.''

\subsubsection{Spatial diffusion}

In presence of a shock, the typical way of computing the spatial diffusion coefficients, $\kappa_\text{x,y,z}$, must be altered to take the ambient gas flow velocity into account. Originally, these coefficients are obtained from the particles' mean square displacements via
\begin{equation}
\kappa_{i,\text{orig}} = \frac{1}{2 t_\text{max}} \left \langle \bigl( x_i(t_\text{max}) - x_i(0) \bigr)^2 \right \rangle, \quad i \in \{ \text{x, y, z} \}.
\end{equation}
In the shock rest frame, the gas flow upstream and downstream of the shock front would also be included in the equation above. This is undesired, as one is rather interested in the diffusion perpendicular or parallel to the magnetic field lines, which, in the flux-freezing approximation, flow together with the ambient plasma. To account for that, time-dependent ``running'' diffusion coefficients are used \citep{tau10:sub}, which are flow-corrected and read
\begin{equation}\label{eq:model-kappa}
\kappa_{i}(t) = \frac{1}{2t} \left \langle \bigl( x_i(t) - x_i(0) - \varDelta x_{\text{flow},i}(t) \bigr)^2 \right \rangle.
\end{equation}
Here, the correction factor $\varDelta \vec{x}_{\text{flow}}(t)$ is introduced, which corresponds to the displacement of the initial particle location $\vec{x}_0$ due to the ambient gas flow. It is defined as
\begin{equation}
\varDelta \vec{x}_\text{flow}(t) = \int_0^t { \vec{R}_\text{flow}(\vec{x}_0(t'))\;\text{d} t'} ,
\end{equation}
where $\vec{R}_\text{flow}(\vec{x_0}(t))$ is either the upstream or downstream flow velocity, depending on the relative position of $\vec{x_0}(t)$ to the shock front. In the \padian simulation, this factor can only be approximated by a discrete summation over a number $N$ of time steps
\begin{equation}
\varDelta \vec{x}_\text{flow}(t) = \sum\limits_{n=0}^N { \vec{R}_\text{flow}\bigl(x_0(n\,\varDelta t)\bigr) \varDelta t} ,
\end{equation}
where $N \varDelta t = t$. It is expected that the impact on the final values of the diffusion coefficient is negligible because the flow velocity is small compared to the particle velocity.

\subsection{Injection}

The work presented here investigates two methods of particle injection. In the simplest case, all particles of the simulation are injected with a constant absolute rigidity value. The second injection profile is inspired by the work of \citet{caprioli2014simulations} and follows the thermal leakage model outlined by \citet{malkov2001modern}. There, the particles are initially assigned an absolute non-relativistic rigidity value following an offsetted Maxwell-Boltzmann distribution of the form
\begin{equation}\label{eq:model-maxwell}
R = R_\text{offs} + f_\text{Gamma} \left (R; 3 / 2, \beta \right) .
\end{equation}
Here, a library function is used for the gamma distribution. Its probability density function in terms of the shape factor $\alpha$ and rate factor $\beta$ is given by
\begin{equation}
f_\text{Gamma}(x; \alpha, \beta) = \frac{\beta^\alpha}{\varGamma(\alpha)} x^{\alpha - 1} \text{e}^{-\beta x},
\end{equation}
where $\varGamma$ denotes the gamma function. For a shape factor $\alpha = 3/2$ this yields $\varGamma(\alpha) = \sqrt{\pi} / 2$. Substituting $\beta = 1/k_\text{B}T$, one then obtains the Maxwell-Boltzmann distribution function of energy
\begin{equation}
f_\text{MW}(E) = 2 \sqrt{\frac{E}{\pi}} \left( \frac{1}{k_\text{B}T} \right)^{3/2} \text{e}^{-E / k_\text{B}T}.
\end{equation}
The rate parameter $\beta$ is configurable in the simulations and defines the width of the distribution. The offset value $R_\text{offs}$ introduced in Eq.~\eqref{eq:model-maxwell} moves the whole distribution into a user defined area.

\begin{figure}
\centering
\includegraphics[bb=235 645 430 775,clip,width=\figwidth]{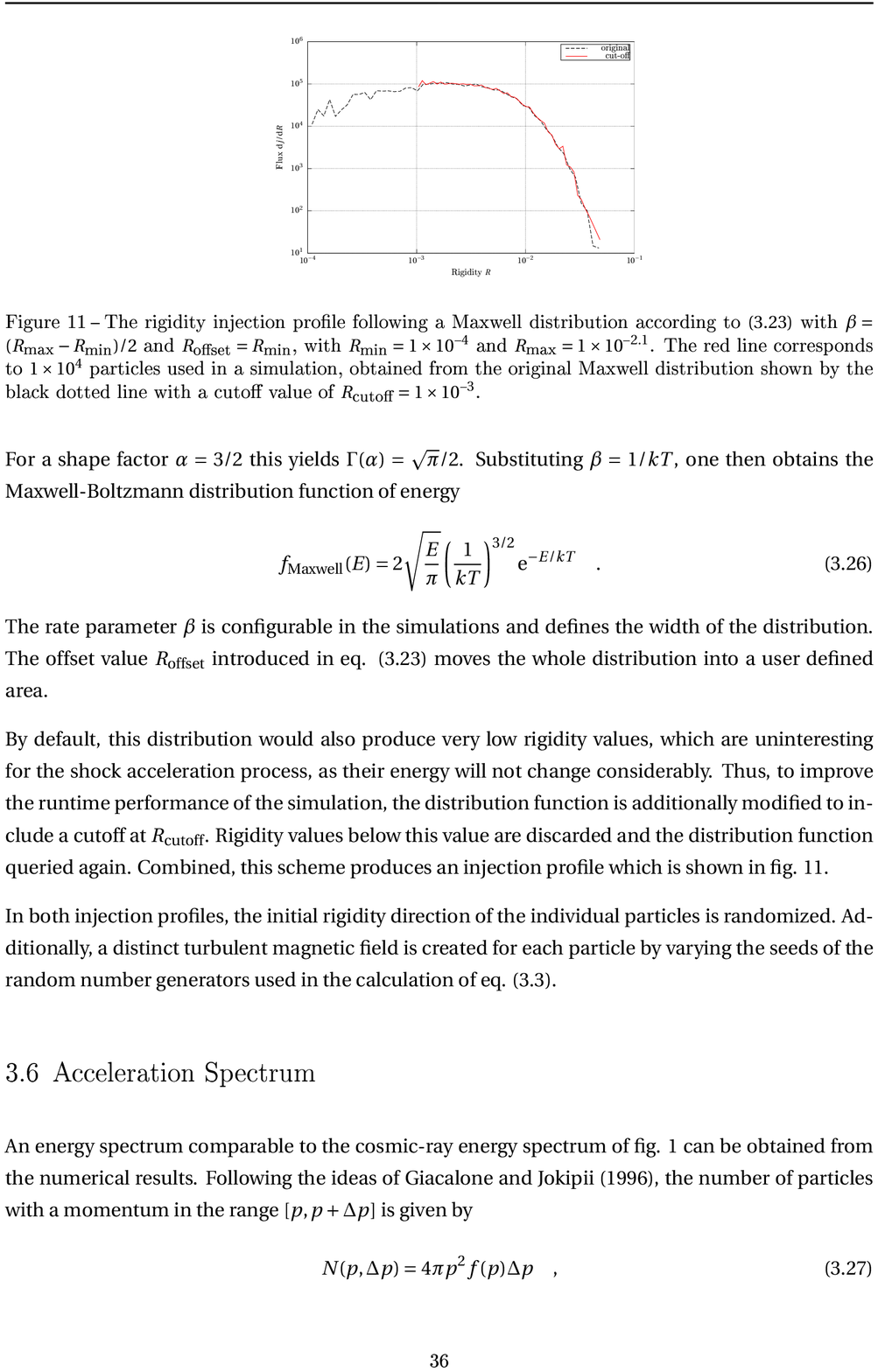}
\caption{Rigidity injection profile following a Maxwell distribution according to \eqref{eq:model-maxwell} with $\beta = (R_\text{max} - R_\text{min})/2$ and $R_\text{offs} = R_\text{min}$, with $R_\text{min} = \SI{E-4}{}$ and $R_\text{max} = \SI{E-2.1}{}$. The red line corresponds to $\SI{E4}{}$ particles used in a simulation, obtained from the original Maxwell distribution shown by the black dotted line with a cutoff value of $R_\text{cutoff} = \SI{E-3}{}$.}
\label{fig:results-maxwell}
\end{figure}

By default, this distribution would also produce very low rigidity values, which are uninteresting for the shock acceleration process, as their energy will not change considerably. Thus, to improve the runtime performance of the simulation, the distribution function is additionally modified to include a cutoff at $R_\text{cutoff}$. Rigidity values below this value are discarded and the distribution function queried again. Combined, this scheme produces an injection profile which is shown in Fig.~\ref{fig:results-maxwell}.

In both injection profiles, the initial rigidity direction of the individual particles is randomized. Additionally, a distinct turbulent magnetic field is created for each particle by varying the seeds of the random number generators used in the calculation of Eq.~\eqref{eq:model-magnetic-field}.

\subsection{Acceleration spectrum}\label{sec:model-powerlaw}

An energy spectrum comparable to the measured cosmic-ray energy spectrum can be obtained from the numerical results as follows. Following the ideas of \citet{giacalone1996perpendicular}, the number of particles with a momentum in the range $[p, p + \varDelta p]$ is given by
\begin{equation}
N(p, \varDelta p) = 4 \pi p^2 f(p) \varDelta p ,
\end{equation}
where $f(p)$ represents the distribution function for the momentum $p$. In terms of the flux $\diff j/\diff E$ this can then be rewritten to yield
\begin{equation}
4 \pi \varDelta E \frac{\diff j}{\diff E} = N(E, \varDelta E) v ,
\end{equation}
where $v$ is the particle's velocity. The above equation can be transformed to use the normalized particle rigidity $R$ applied by \padian, instead of the energy. Starting with the energy
\begin{equation}
E = \gamma m \text{c}^2 ,
\end{equation}
one can substitute the Lorentz factor $\gamma = \sqrt{1 + (p / m \text{c})^2}$ yielding
\begin{equation}
E = \sqrt{\left(m \c^2 \right)^2 + \left(p \c\right)^2} .
\end{equation}
Replacing the momentum with $p = R \varOmega m l_0$ and simplifying the expression leads to
\begin{equation}
E = \varOmega m l_0 \c \sqrt{R^2 + R_\c^2} ,
\end{equation}
with $R_\c = c / \varOmega l_0$. Under the assumption of $R \gg R_\text{c}$, which is valid as long as $v/c \gg 1/\sqrt{2}$, one arrives at $E \propto R$. The equation for the power spectrum can thus be written as
\begin{equation}\label{eq:model-powerlaw}
\frac{\diff j}{\diff E} \propto \frac{N(R, \varDelta R)}{R\,\varDelta R}.
\end{equation}
In the non-relativistic limit $R \ll R_\text{c}$ on the other hand, $E_\text{kin} \propto R^2$, leading to
\begin{equation}\label{eq:model-powerlaw-kin}
\frac{\diff j}{\diff E_\text{kin}} \propto \frac{N(R^2, \varDelta R^2)}{R^2 \varDelta R^2} .
\end{equation}
In the spectra plots below, logarithmic spacing is chosen by sampling the above equation with a constant ratio $\varDelta R/R$.

\section{Simulation results}\label{results}

The following chapter presents various results obtained from running \padian under different parameter configurations.
To reduce the parameter space, the following values where set in all simulations:

\begin{itemize}
\item The compression ratio $r$ is set to $4$.
\item The magnetic mean field $B_0$ is set to be uniform and points along the $z$-axis.
\item Two types of turbulent magnetic field $\delta B$ generation are studied. On one hand, the magnetostatic case according to Eq.~\eqref{eq:model-magnetic-field}, or via \alfven waves and Eq.~\eqref{eq:model-magnetic-field-alfven}. In both cases, $N_\text{m} = 128$ wave modes are superposed in slab geometry. A Kolmogorov energy spectrum is chosen for the magnetic turbulence, i.\,e. $s = 5/3$, with an additional energy range proportional to $k^q$, here with $q = 0$. The minimum wavenumber exponent in units of the bend-over scale is set to $-5$ and the maximum one to $3$. Finally, the turbulent magnetic field is normalized to fulfill the condition $\delta B / B_0 = 1$, corresponding to a model of strong turbulence.
\item For every particle injected into the simulation, a separate turbulent field is created with different random seed values. This ensures the independence of individual test particles when analyzing all as a statistical ensemble.
\end{itemize}

\begin{figure}
\centering
\includegraphics[bb=72 610 260 735,clip,width=\figwidth]{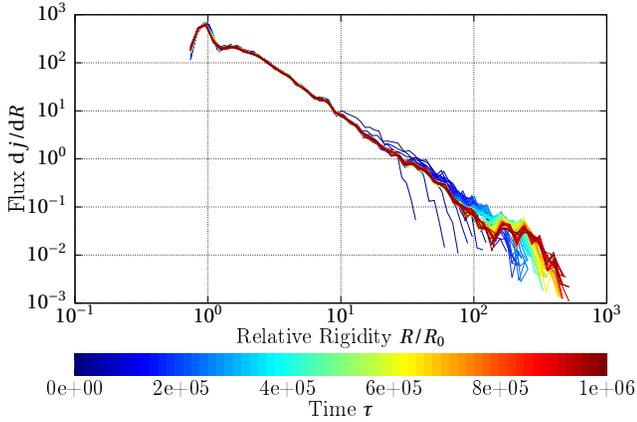}
\caption{Parallel magnetostatic shock with particles injected at a fixed rigidity of $R(0)=10^{-2}$. The time evolution of the power spectrum, i.\,e., the flux $\diff j/\diff R$ over total rigidity gain $R/R(0)$, over the runtime of a \padian simulation of a strong parallel shock with $10^4$ test particles is shown. The feature on the left side indicates that some particles lose energy. To the right, a smooth constant power law establishes itself over time.}
\label{fig:results-parallel-time}
\end{figure}

\begin{figure}
\centering
\includegraphics[bb=280 610 465 735,clip,width=\figwidth]{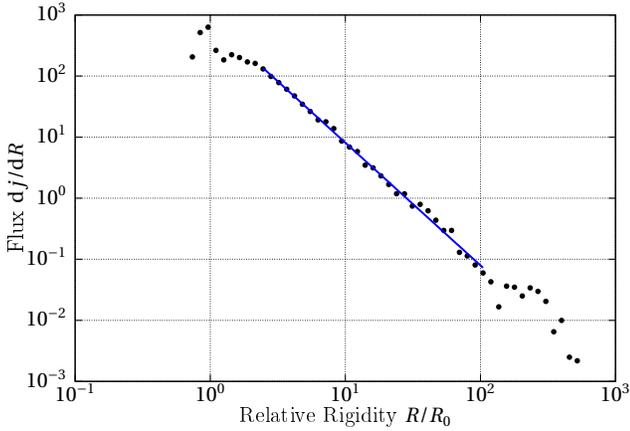}
\caption{Same as Fig.~\ref{fig:results-parallel-time} but only for the last time step. A linear fit of the double logarithmic data of the final power spectrum for the last time step at $\tau = \SI{E6}{}$ is shown. For the data range spanned by the blue fit line, a spectral index $x \approx -2.01 \pm 0.03$ is obtained.}
\label{fig:results-parallel-fit}
\end{figure}

For all simulations, the particle energy spectrum at different times is obtained. This allows one to estimate: (i) the influence of the initial particle energy spectrum; (ii) the method of magnetic field turbulence generation; and (iii) the direction of the shock has on the final spectrum. Within the resulting particle energy spectra, a range is selected and its double-logarithmic data fitted linearly to find the spectral index. Note that the errors given for the spectral index come from this analysis alone. While all results are qualitatively reproducible, differently seeded \padian simulations with otherwise equal configuration parameters would yield slightly different numerical results in the order of the fit errors.

\subsection{Benchmark results}\label{sec:results-magnetostatic-parallel}

Initially, different configurations of a system with magnetostatic turbulence are investigated. The turbulent magnetic field component is generated using Eq.~\eqref{eq:model-magnetic-field} and can easily be adapted to oblique shocks.

First, the effects of shocks on an ensemble of $\SI{E4}{}$ particles injected at the origin with an equal rigidity value of $R = \SI{E-2}{}$ is investigated. All particles are assigned random initial rigidity directions. The shock rigidity is set to $R_S = R / \sqrt{3} \approx \SI{5.8E-3}{}$, following the previous work by \citet{giacalone1999transport}. The simulation runs in the shock rest frame, where the shock front lies in a plane through the origin, such that all particles are injected directly at the shock. This system is then simulated until the normalized time $\tau = \varOmega_{\text{rel}} t=\SI{E6}{}$ is reached with $\varOmega_{\text{rel}}$ the relativistic gyrofrequency. At 100 equidistant time points, the power spectrum of all particles is computed according to Eq.~\eqref{eq:model-powerlaw}.

For a parallel shock, where the shock front lies in the $x$-$y$-plane, the results are shown in Fig.~\ref{fig:results-parallel-time}. One can observe multiple effects: First, note how a fraction of the particles lose energy. This effect is a result of the relative motion of the ambient gas on the rigidity of the injected particles and unrelated to the shock acceleration. Second, all other particles are efficiently accelerated by the shock. Already after the first snapshot, a power law is exhibited for parts of the energy spectrum. Over time, this range grows as particles continue to be accelerated. A linear fit of the double-logarithmic power spectrum data in this range, as is shown in Fig.~\ref{fig:results-parallel-fit}, yields a spectral index of about $x \approx \SI{-2.01 \pm 0.03}{}$. This value is compatible with the theoretically predicated value of $-2$.

\begin{figure}
\centering
\includegraphics[bb=68 638 260 765,clip,width=\figwidth]{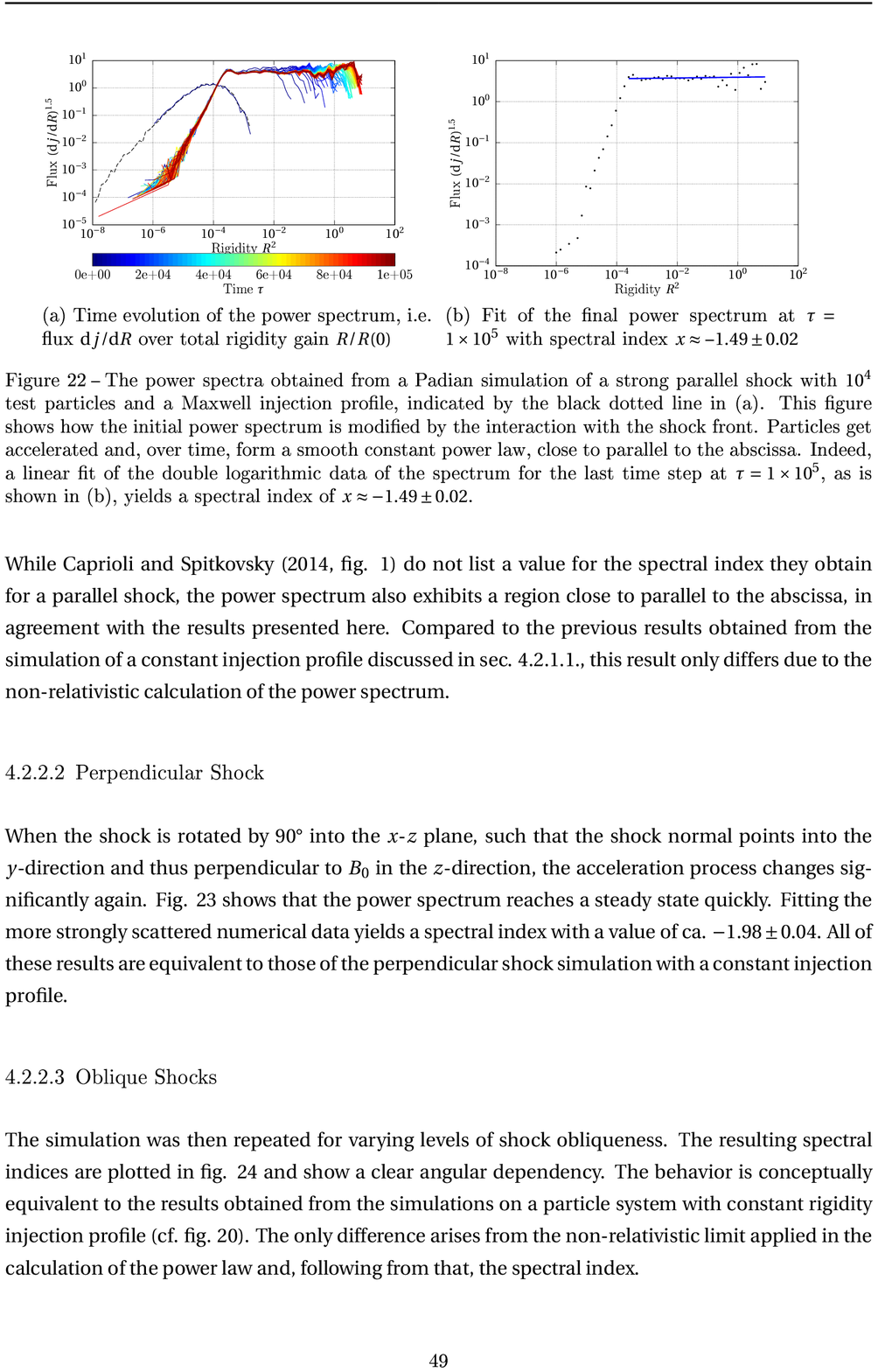}
\caption{Parallel magnetostatic shock with a Maxwellian injection profile as indicated by the black dotted line. This figure shows how the initial power spectrum is modified by the interaction with the shock front. Particles get accelerated and, over time, form a smooth constant power law, close to parallel to the abscissa.}
\label{fig:results-maxwell-parallel-time}
\end{figure}

\begin{figure}
\centering
\includegraphics[bb=275 638 464 765,clip,width=\figwidth]{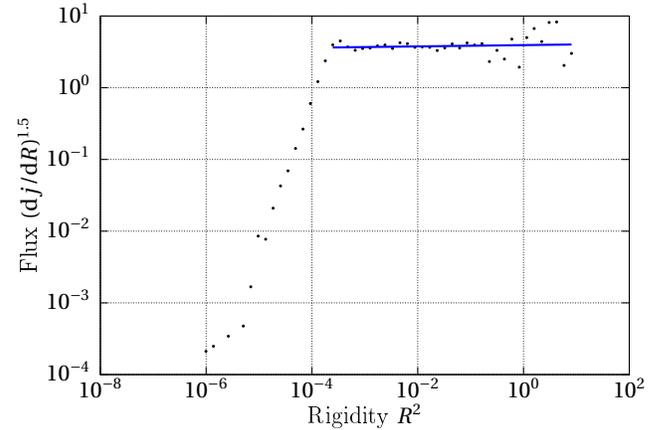}
\caption{Same as Fig.~\ref{fig:results-maxwell-parallel-time} but only for the last time step. A linear fit of the double logarithmic data of the spectrum for the last time step at $\tau = \SI{E5}{}$ yields a spectral index of $x \approx -1.49 \pm 0.02$.}
\label{fig:results-maxwell-parallel-fit}
\end{figure}

Inspection of the individual particles' relative rigidity boost over the shock distance shows some interesting insights. First, some particles are still close to the shock front and thus within the acceleration zone, which explains the continuous rise of the total energy of the particle ensemble. Second, it appears that the bulk of particles gets advected downstream of the shock. These particles have a distance close to $-R_\text{S} \tau / r$, indicating that, under the flux-freezing assumption, they follow the magnetic field lines which flow with the downstream ambient plasma velocity of $R_\text{S} / r$.

The total kinetic energy of the system increases by a factor of close to $7$ over the simulation time. The total aggregated rigidity of all particles still rises toward the end of the simulation. This is an indication that the simulation did not yet reach a steady state. By increasing the simulation time further, even higher maximum particle energies are to be expected. However, no qualitative changes to the power law are expected, just the high-energy tail will become smoother.

\subsection{Maxwellian injection profile}

\begin{figure}
\centering
\includegraphics[bb=118 638 310 766,clip,width=\figwidth]{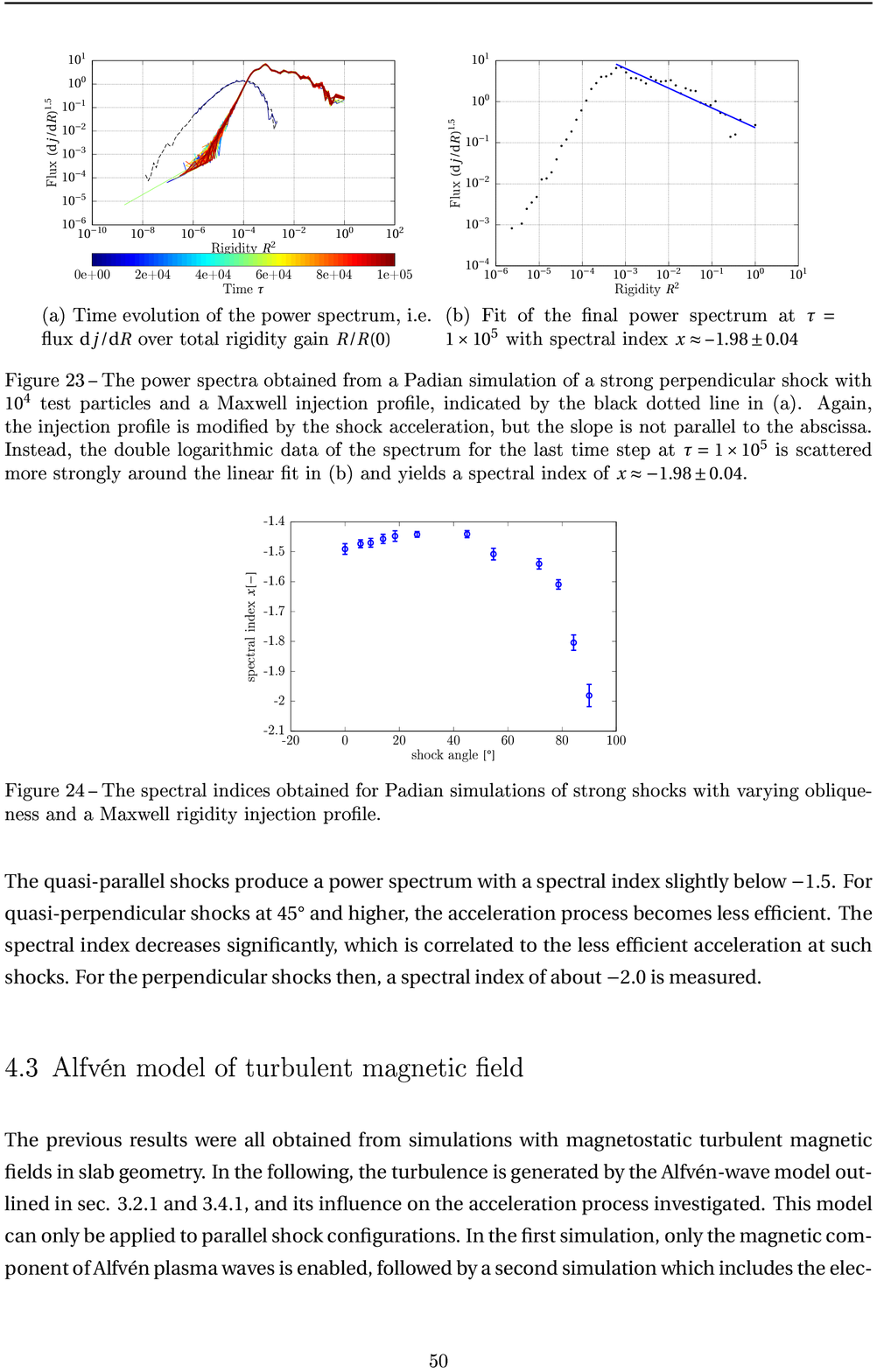}
\caption{Perpendicular magnetostatic shock with a Maxwellian injection profile as indicated by the black dotted line. Again, the injection profile is modified by the shock acceleration, but the slope is not parallel to the abscissa.}
\label{fig:results-maxwell-perpendicular-time}
\end{figure}

\begin{figure}
\centering
\includegraphics[bb=324 638 514 766,clip,width=\figwidth]{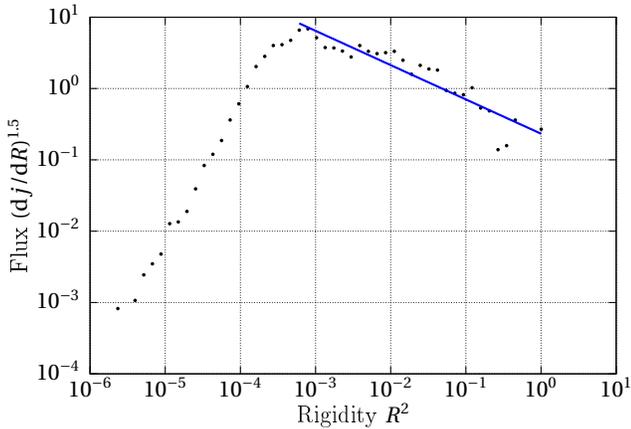}
\caption{Same as Fig.~\ref{fig:results-maxwell-perpendicular-time} but only for the last time step. The double logarithmic data of the spectrum for the last time step at $\tau = \SI{E5}{}$ is scattered more strongly around the linear fit and yields a spectral index of $x \approx -1.98 \pm 0.04$.}
\label{fig:results-maxwell-perpendicular-fit}
\end{figure}

Inspired by the work of \citet{caprioli2014simulations}, the effect of the shock acceleration on a particle ensemble with an initially Maxwell-distributed rigidity spectrum is also investigated. The simulation was adapted such that each of the \SI{E4}{} particles is injected with an absolute rigidity value given by Eq.~\eqref{eq:model-maxwell}. For the simulations below, the distribution parameters were set to $\beta = (R_\text{max} - R_\text{min})/2$ and $R_\text{offs} = R_\text{min}$, with $R_\text{min} = \SI{E-4}{}$ and $R_\text{max} = \SI{E-2.1}{}$. The rigidity cutoff was set to $R_\text{cutoff} = \SI{E-3}{}$. The initial direction of each particle is, as before, still random. The shock rigidity is set to $R_\text{S} = \SI{E-2}{}$ and only a limited time period up to $\tau = \SI{E5}{}$ is simulated. Increasing this value improves the smoothness of the power spectrum in the high-rigidity regime, at the cost of significantly longer simulation run times, but does not influence the results for the power spectra qualitatively.

For comparison purposes with the original work of \citet{caprioli2014simulations}, the power spectra are now obtained by plotting the flux to the power of $1.5$ over the square of the rigidity. This relation corresponds to a particles' kinetic energy dependence on the rigidity in the non-relativistic limit (cf. Sec.~\ref{sec:model-powerlaw}). This is necessary because the theory of the Maxwell-Boltzman distribution is only applicable to non-relativistic energies. Contrary to the previous injection model with relativistic constant rigidity, one now expects spectral indices of $1.5$ from theory \citep{caprioli2014simulations}.

Furthermore, one cannot directly calculate diffusion coefficients for this model of injection. The formula in Eq.~\eqref{eq:model-kappa} is only meaningful for particles of similar starting energy. Here, one would need to group the injected particles by energy and calculate individual coefficients, which requires a significant increase in the number of injected particles. Due to the higher computational effort this would require, the spatial diffusion coefficients are omitted below.

\subsubsection{Parallel shock}\label{sec:results-maxwell-parallel}

For a parallel shock configuration a smooth power spectrum with a spectral index of about $-1.5$ quickly establishes itself, as can be seen in Fig.~\ref{fig:results-maxwell-parallel-time}. The high-rigidity tail of the power spectrum is still in flux and expected to smooth out for longer simulation times.

It is notable how the injection profile is influenced also in the low-rigidity region below the shock rigidity. This effect might be due to the cutoff applied to the injection profile. However, while removing this may influence the shape of the power spectra, no qualitative change to the spectral index in the constant power law region is expected. Rather, one can expect that the direct injection at the shock front leads to this result. If significantly more particles are injected, in a larger space around the shock front, the low-energy tail of the power spectrum should stay rather unaffected. But, since this requires significantly higher computational effort, this assumption was not verified in the work presented here.

While \citet[fig. 1]{caprioli2014simulations} do not list a value for the spectral index they obtain for a parallel shock, the power spectrum also exhibits a region close to parallel to the abscissa, in agreement with the results presented here. Compared to the previous results obtained from the simulation of a constant injection profile discussed in Sec.~\ref{sec:results-magnetostatic-parallel}, this result only differs due to the non-relativistic calculation of the power spectrum.

\subsubsection{Perpendicular shock}

\begin{figure}
\centering
\includegraphics[bb=128 646 324 775,clip,width=\figwidth]{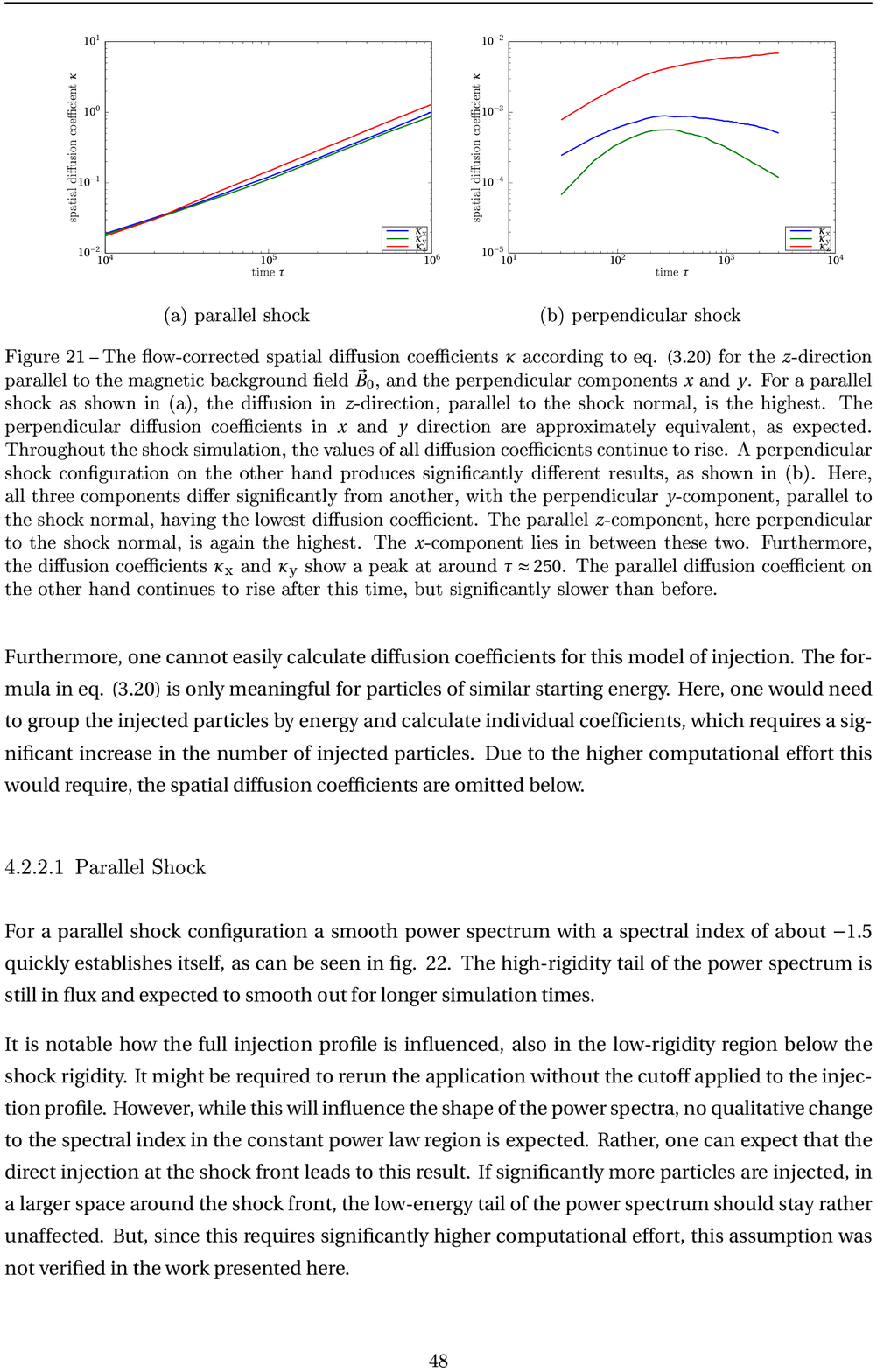}\\[5pt]
\includegraphics[bb=336 646 532 775,clip,width=\figwidth]{kappa_time}
\caption{Flow-corrected spatial diffusion coefficients $\kappa$ according to Eq.~\eqref{eq:model-kappa} for the $z$-direction parallel to the magnetic background field $\vec{B}_0$, and the perpendicular components $x$ and $y$. For a parallel shock as shown in the upper panel, the diffusion in $z$-direction, parallel to the shock normal, is the highest. The perpendicular diffusion coefficients in $x$ and $y$ direction are approximately equivalent, as expected. Throughout the shock simulation, the values of all diffusion coefficients continue to rise. A perpendicular shock configuration on the other hand produces significantly different results, as shown in the lower panel. Here, all three components differ significantly from another, with the perpendicular $y$-component, parallel to the shock normal, having the lowest diffusion coefficient. The parallel $z$-component, here perpendicular to the shock normal, is again the highest. The $x$-component lies in between these two. Furthermore, the diffusion 
coefficients $\kappa_\text{x}$ and $\kappa_\text{y}$ show a peak at around $\tau \approx 250$. The parallel diffusion coefficient on the other hand continues to rise after this time, but significantly slower than before.}
\label{fig:results-kappa}
\end{figure}

For a perpendicular shock, the simulation was set up as before in Sec.~\ref{sec:results-maxwell-parallel}, but the shock front now lies in the $x$-$z$-plane. The background magnetic field $\vec{B}_0$ still points in the $z$-direction, and is thus transverse to the shock normal. Downstream of the shock, the transverse components of the combined magnetic field $\vec{B}$ are amplified by the shock compression ratio. In this case, the results differ strongly from those of a parallel shock. Most notably, the particles are only accelerated during a short time period, as seen in Fig.~\ref{fig:results-maxwell-perpendicular-time}.
\hl{This is in agreement with previous studies \citep[and references therein]{gia05:sho,caprioli2014simulations}, who showed that perpendicular shocks accelerate particles at a higher rate than parallel shocks. In particular, \citet{gia05:sho} found that the flux at the highest energies is dominated by particles accelerated at a perpendicular shock.}

The particles diffusion coefficient in these directions is significantly larger than in the normal, $y$-direction, as shown in Fig.~\ref{fig:results-kappa} for a constant injection profile. Once the particles cross the shock front, they are advected from the acceleration zone with the ambient downstream plasma flow. Because of that, a much shorter simulation time of only $\tau = \SI{3E3}{}$ is sufficient to capture the dynamics of the acceleration process, indicated also by the plateau reached in the total rigidity. Within this time, the diffusion coefficients leave the initial ballistic regime (see Fig.~\ref{fig:results-kappa}). After reaching a peak at around $\tau = 250$, the transverse diffusion coefficients $\kappa_{\text{x,y}}$ decrease over time. Due to $\langle B_\text{x,y} \rangle = 0$ one expects these values to approach zero for even higher times. The diffusion along $z$, i.\,e. parallel to the magnetic mean field but transverse to the shock normal, also decreases its growth at this time, but still increases slightly. With $\left \langle B_\text{z}  \right \rangle = B_0$, one can expect $\kappa_\text{z}$ to approach a steady state for higher simulation times. The strong initial increase of the diffusion coefficients, especially in the $y$-direction, correlates to the acceleration process: with increased energy, the Larmor radii of the particles increases and thus the diffusion coefficients increase.

The resulting power spectrum for this shock configuration is shown in Fig.~\ref{fig:results-maxwell-perpendicular-fit}. Fitting the more strongly scattered numerical data yields a spectral index with a value of $-1.98 \pm 0.04$ corresponding to a constant power law. Note that this result is equivalent to that of a perpendicular shock with a constant injection profile. One can still observe efficient acceleration during the short acceleration period, with individual particle rigidities being boosted by about two orders of magnitude. Overall, the combined system increases its kinetic energy by a factor of close to $2.5$.

\subsubsection{Oblique shocks}

\begin{figure}
\centering
\includegraphics[bb=220 398 420 528,clip,width=\figwidth]{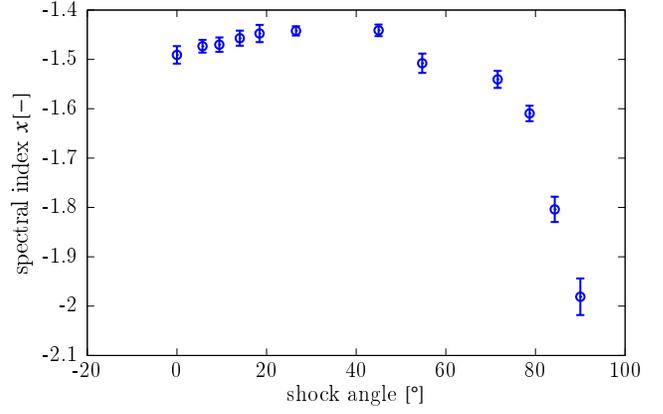}
\caption{Oblique magnetostatic shock with a Maxwellian injection profile. The spectral indices obtained for \padian simulations of strong shocks with varying obliqueness and a Maxwell rigidity injection profile.}
\label{fig:results-maxwell-oblique}
\end{figure}

\begin{figure}
\centering
\includegraphics[bb=274 640 474 770,clip,width=\figwidth]{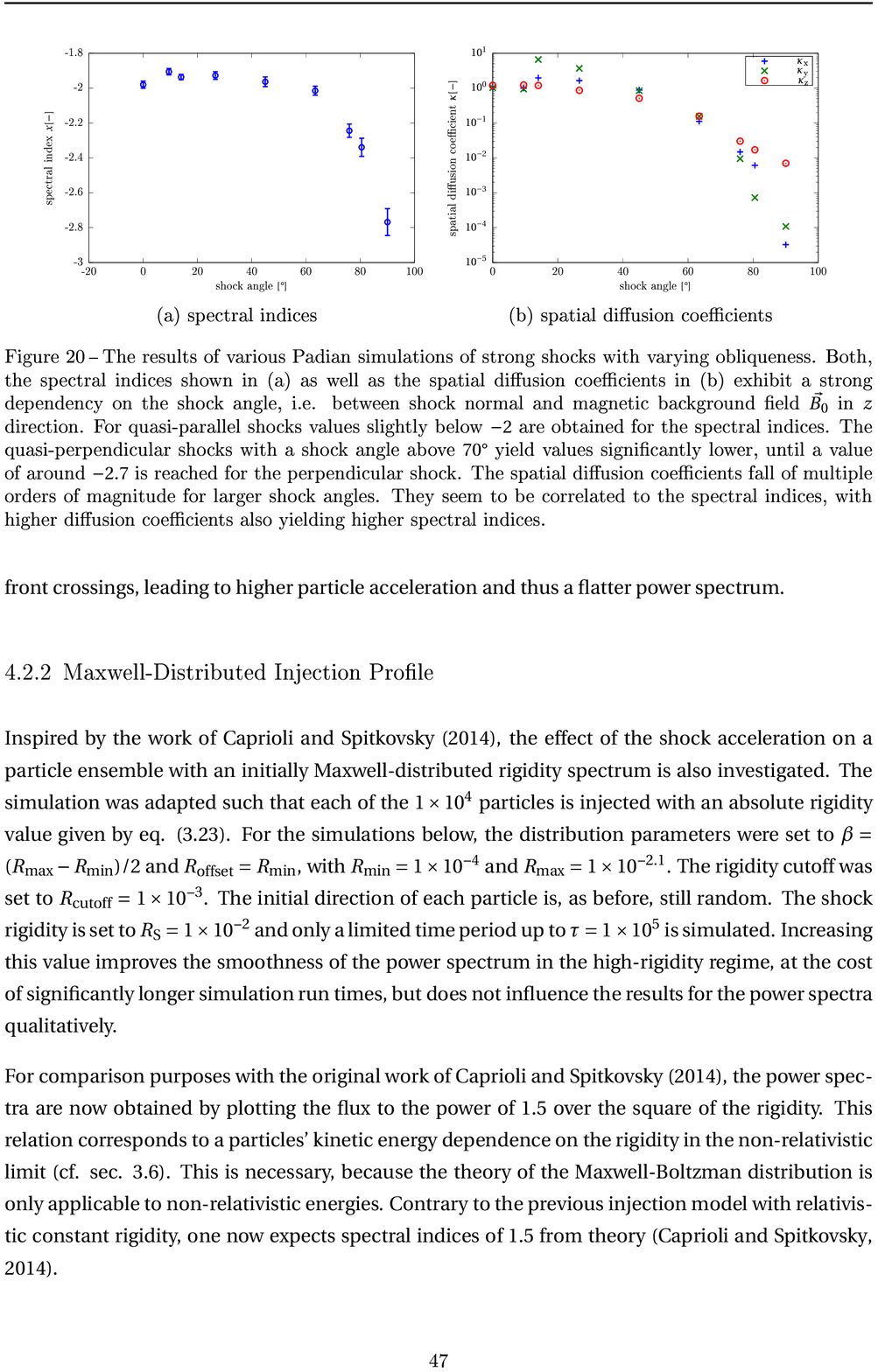}
\caption{Spatial diffusion coefficients for various shock obliqueness angles, showing a variation by multiple orders of magnitude for larger shock angles. They seem to be correlated to the spectral indices (cf. Fig.~\ref{fig:results-maxwell-oblique}, with higher diffusion coefficients also yielding higher spectral indices. Note that for the calculation of diffusion coefficients, a constant injection profile has been assumed.}
\label{fig:results-oblique-kappa}
\end{figure}

Between the two extreme cases of a parallel or perpendicular shock, the so called oblique shocks can be set up \citep{sir09:obl}. For simulations using a varying level of the shock obliqueness, the resulting spectral indices are plotted in Fig.~\ref{fig:results-maxwell-oblique}, showing a clear angular dependency. Again, the behavior is qualitatively equivalent to the results obtained from the simulations on a particle system with a constant rigidity injection profile.

The quasi-parallel shocks produce a power spectrum with a spectral index slightly below $-1.5$. For quasi-perpendicular shocks at $\SI{45}{\degree}$ and higher, the acceleration process becomes less efficient. The spectral index decreases significantly, which is correlated to the less efficient acceleration at such shocks. For the perpendicular shocks then, a spectral index of about $-2.0$ is measured.

\begin{figure}
\centering
\includegraphics[bb=122 598 310 724,clip,width=\figwidth]{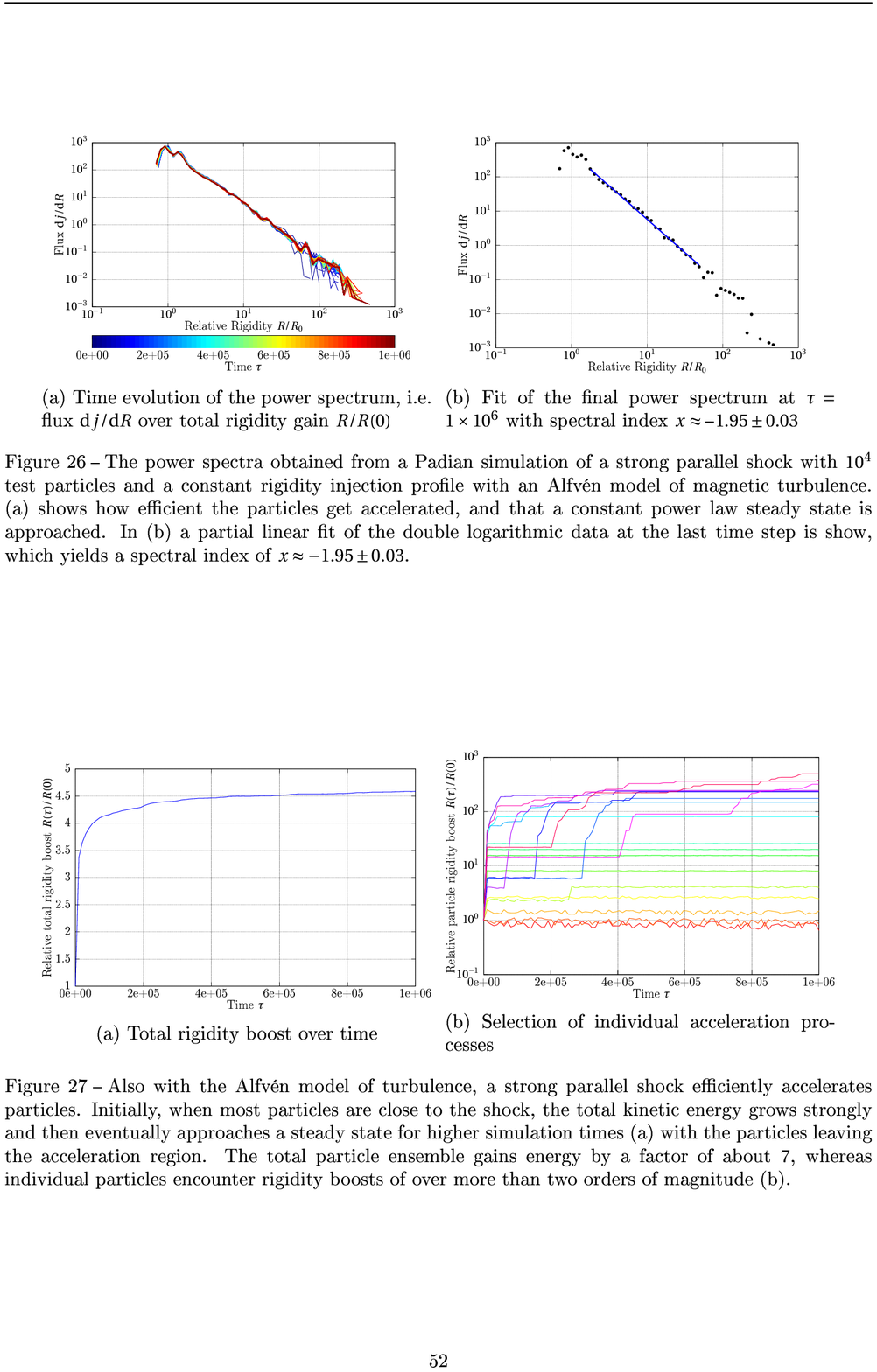}
\caption{Parallel \alfven{}ic shock with particles injected at a fixed rigidity of $R(0)=10^{-2}$. The time evolution of the power spectrum shows shows how efficient the particles get accelerated, and that a constant power law steady state is approached.}
\label{fig:results-wave-parallel-time}
\end{figure}

\begin{figure}
\centering
\includegraphics[bb=330 598 514 724,clip,width=\figwidth]{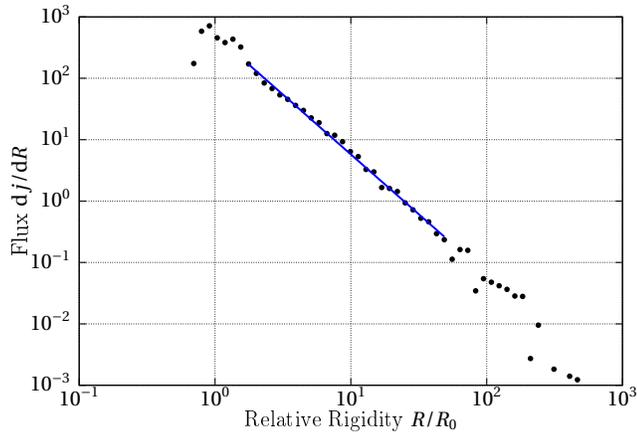}
\caption{Same as Fig.~\ref{fig:results-wave-parallel-time} but only for the last time step. A (partially) linear fit of the double logarithmic data at the last time step is shown, which yields a spectral index of $x \approx -1.95 \pm 0.03$.}
\label{fig:results-wave-parallel-fit}
\end{figure}

\begin{figure*}
\centering
\includegraphics[bb=116 270 320 401,clip,width=0.47\linewidth]{spectrum_wave_par}\quad
\includegraphics[bb=116 441 320 578,clip,width=0.47\linewidth]{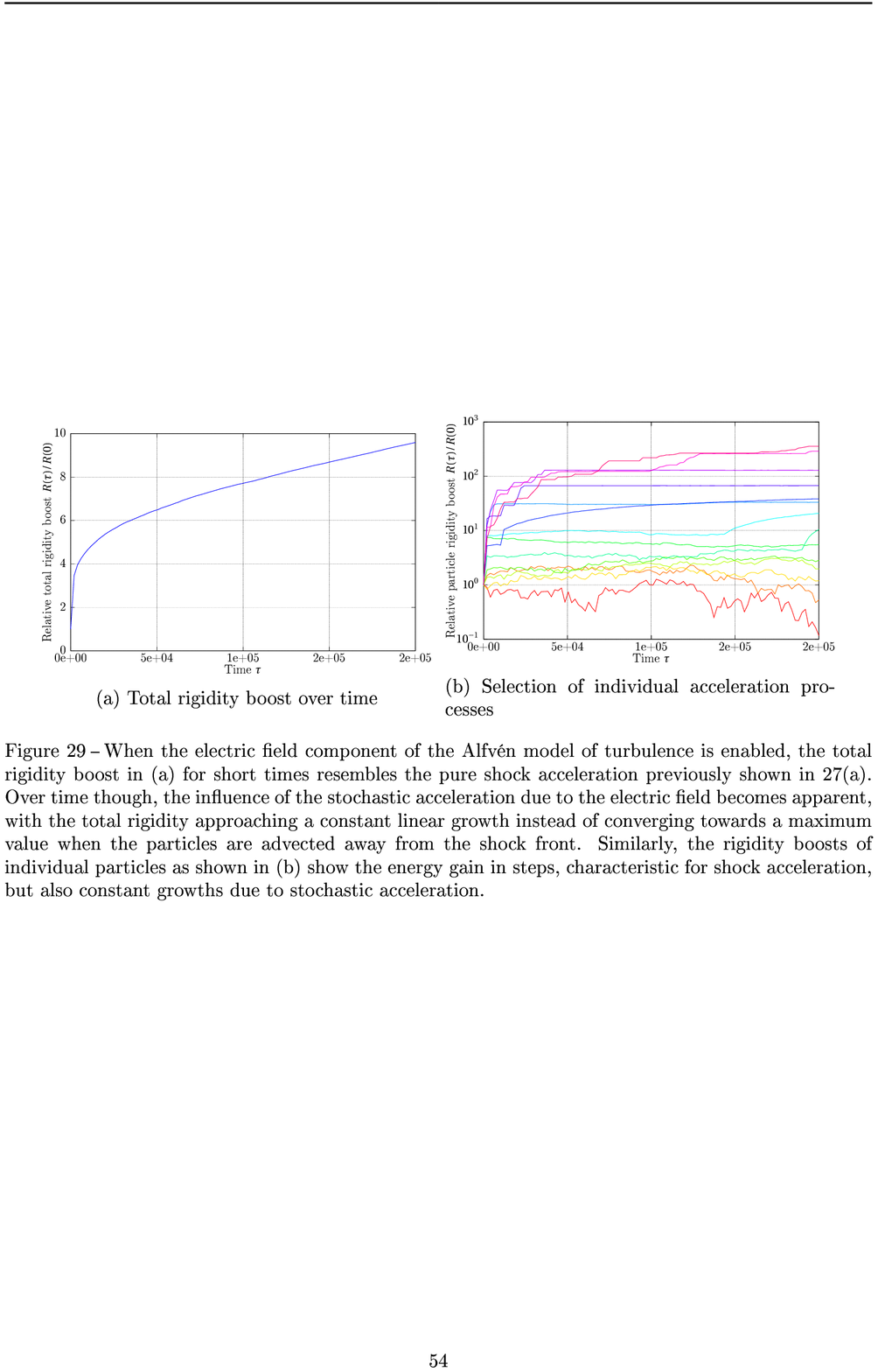}\\[5pt]
\includegraphics[bb=324 275 528 407,clip,width=0.47\linewidth]{spectrum_wave_par}\quad
\includegraphics[bb=324 448 528 580,clip,width=0.47\linewidth]{indiv_wave_electric}
\caption{Parallel \alfven{}ic shock with particles injected at a fixed rigidity of $R(0)=10^{-2}$. Shown are the cases without (left panels) and with (right panels) turbulent electric fields. As shown in the upper panels, a strong shock efficiently accelerates the particle ensemble. The acceleration is strongest at the beginning, when all particles are close to the shock front. Due to parallel diffusion, many particles will not encounter the shock again after the initial boost, whereas others continue to get accelerated. This is apparent from the lower panels, which also show that some particles initially lose energy or never cross the shock to get accelerated.}
\label{fig:results-parallel-kin}
\end{figure*}

\hl{Our results are, to some extend, corroborated by the findings of \citet[their Fig.~12]{caprioli2014simulations}, although their explanation relies on the central fact that the back-reaction of the particles on the shock structure is included. Accordingly, the structure of the self-generated turbulent magnetic fields in the upstream regime depends strongly on the shock angle. In contrast, \citet{gia05:sho} found that, in the high-energy end of the particle spectrum, the spectral index tends to be compatible with the predictions of diffusion shock acceleration model, independently of the shock normal angle. However, it should be noted that they obtained the high-energy tail of the spectrum with the help of a special steady-state method. At intermediate energies, the spectral index becomes steeper with an increasing shock angle, similar to our findings.}

\hl{A possible explanation for our results can be found by modifying the core assumptions of diffusive shock acceleration. In particular, the escape probability might be related to the diffusive behavior of the particles and, thus, might be energy-dependent. To illustrate this further, note that, to lowest order,}
particles can be assumed to follow the magnetic field lines, which, on average, are given by the magnetic mean field $B_0$. The diffusion coefficient parallel to the direction of $B_0$, $\kappa_\parallel$ is significantly higher than that in the perpendicular direction, $\kappa_\perp$, as shown in Fig.~\ref{fig:results-kappa}. In addition, the diffusion coefficients also have a different energy dependence. In quasi-perpendicular shocks, the magnetic field lines are advected with the downstream plasma flow, and the particles will then follow this flow, with $\kappa_\text{y}$ being the smallest of the three diffusion coefficient components. Additionally, the increased magnetic field strength downstream reduces the particles' Larmor radius. Thus the acceleration process only occurs for a short time period, and just a few particles reach high energies, leading to the steep power law. The results for the diffusion coefficients for varying shock obliqueness, as shown in Fig.~\ref{fig:results-oblique-kappa}, also seem to validate this hypothesis. There, a correlation of the spectral index and the diffusion coefficients is visible: the higher diffusion coefficients in the direction parallel to the shock front enable more shock front crossings, leading to higher particle acceleration and thus a flatter power spectrum. \hl{For further discussion, the reader is referred to the investigations of shock drift acceleration \citep[e.\,g.,][and references therein]{bal01:sda,par13:sda}, which underline the complex interactions of particles with quasi-perpendicular shocks.}

\subsection{\alfven{}ic turbulent magnetic field}

The previous results were all obtained from simulations with magnetostatic turbulent magnetic fields in slab geometry. In this section, the turbulence is generated by the \alfven-wave model outlined in Sec.~\ref{sec:model-alfven} and \ref{sec:model-alfven-shock}, and its influence on the acceleration process investigated. This model can only be applied to parallel shock configurations. In the first simulation, only the magnetic component of \alfven plasma waves is enabled, followed by a second simulation which includes the electric field component to investigate the effect of stochastic acceleration. All other parameters for the simulations presented in the following are copied from Sec.~\ref{sec:results-magnetostatic-parallel}. Most notably, particles are injected with a constant rigidity of $\SI{E-2}{}$, and the shock rigidity is set to $\SI{E-2}{} / \sqrt{3}$.

Running \padian with such a configuration yields power spectra as shown in Fig.~\ref{fig:results-wave-parallel-time}, which again indicate a clear acceleration process occurring at the shock front. Compared to the results for magnetostatic turbulence, a slightly lower value of $x \approx \SI{-1.95 \pm 0.03}{}$ is measured for the spectral index of the power spectrum at the last time step (Fig.~\ref{fig:results-wave-parallel-fit}). Otherwise, no fundamental differences are apparent. Likewise, the growth of the total kinetic energy of the particle system, as shown in Fig.~\ref{fig:results-parallel-kin}, resembles the previous data for magnetostatic turbulence that was presented in subsection~\ref{sec:results-maxwell-parallel}.

\subsubsection{Influence of the electric field}

\begin{figure}[t]
\centering
\includegraphics[bb=72 642 260 765,clip,width=\figwidth]{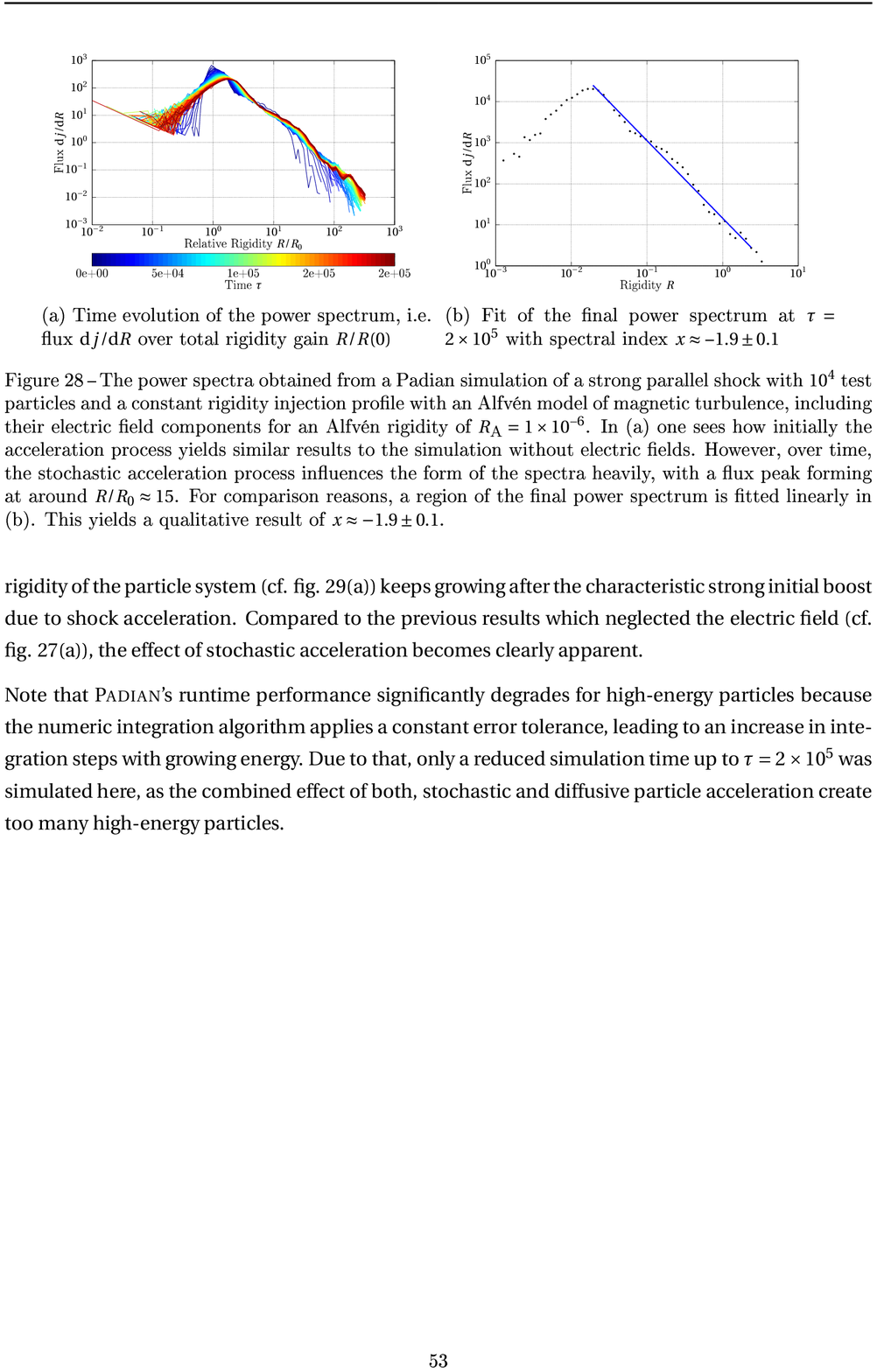}
\caption{Parallel \alfven{}ic shock with particles injected at a fixed rigidity of $R(0)=10^{-2}$. In contrast to Fig.~\ref{fig:results-wave-parallel-time} and~\ref{fig:results-wave-parallel-fit}, the turbulent electric field components for an \alfven rigidity of $R_\text{A} = \SI{E-6}{}$ is also included. The figure shows how initially the acceleration process yields similar results to the simulation without electric fields. However, over time, the stochastic acceleration process influences the form of the spectra heavily, with a flux peak forming at around $R/R_0 \approx 15$.}
\label{fig:results-efield-parallel-time}
\end{figure}

\begin{figure}
\centering
\includegraphics[bb=282 642 464 765,clip,width=\figwidth]{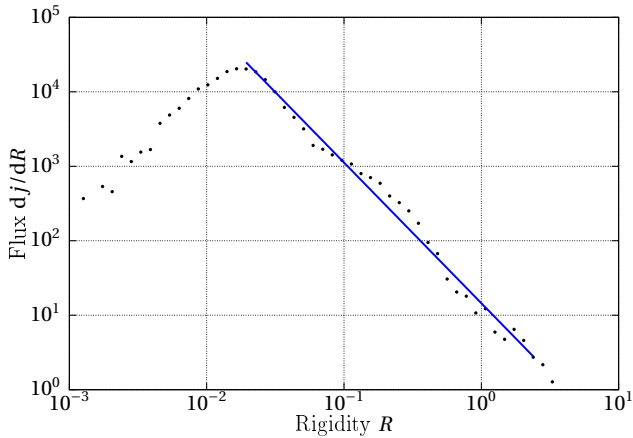}
\caption{Same as Fig.~\ref{fig:results-efield-parallel-time} but only for the last time step. For comparison reasons, a region of the final power spectrum is fitted linearly. This yields a qualitative result of $x \approx -1.9 \pm 0.1$.}
\label{fig:results-efield-parallel-fit}
\end{figure}

When the electric field component of the \alfven plasma waves are included in the simulation, the results are significantly modified. Fig.~\ref{fig:results-efield-parallel-time} shows the power spectra obtained from such a configuration, and Fig.~\ref{fig:results-parallel-kin} depicts the acceleration process in terms of the kinetic energy boost of the total particle system as well as for a selection of individual particles. Compared to the simulation without electric fields, one sees a larger fraction of particles losing energy, indicated by the relatively large flux of particles with relative rigidities below $1$ in the power spectrum. Furthermore, no constant power law is visible, but rather a second peak at around $R/R_0 \approx 15$ grows over time. A fit of the final power spectrum (Fig.~\ref{fig:results-efield-parallel-fit}) to find a spectral index for comparison reasons yields a value of $x \approx -1.9 \pm 0.1$. The total rigidity of the particle system (cf. Fig.~\ref{fig:results-parallel-kin}) keeps growing after the characteristic strong initial boost due to shock acceleration.

Interestingly, the total rigidity boost is of the order of one magnitude, whereas individual particle boost lie in the range of up to three orders of magnitude. The total particle ensemble gains energy by a factor of about $7$, whereas individual particles encounter rigidity boosts of over more than two orders of magnitude. When the electric field component of the \alfven model of turbulence is enabled, the total rigidity boost for short times resembles the pure shock acceleration. Over time though, the influence of the stochastic acceleration due to the electric field becomes apparent, with the total rigidity approaching a constant linear growth instead of converging toward a maximum value when the particles are advected away from the shock front \citep[cf.][]{tau13:sto}.
\hl{In the absence of a shock, the stochastic acceleration due to turbulent electric fields is well known and has been investigated theoretically in terms of momentum diffusion \citep[e.\,g.,][]{ski75:alf,sch89:cr1,sch94:mom} as well as numerically \citep[e.\,g.,][]{mic99:mom,tau10:wav}. Time-dependent turbulence in the form of magnetohydrodynamic plasma waves such as Alfv\'en waves can be interpreted as an acceleration mechanism comparable to a first-order Fermi process \citep{sch09:fer}.}

In combination, the rigidity boosts of individual particles show the energy gain in steps, characteristic for shock acceleration, but also constant growths due to stochastic acceleration. Compared to the previous results which neglected the electric field, the effect of stochastic acceleration are thus clearly apparent.

Note that \padian's runtime performance significantly degrades for high-energy particles because the numeric integration algorithm applies a constant error tolerance, leading to an increase in integration steps with growing energy. Due to that, only a reduced simulation time up to $\tau = \SI{2E5}{}$ was simulated here, as the combined effect of both, stochastic and diffusive particle acceleration create too many high-energy particles.

\section{Discussion and conclusion}\label{summ}

In this article, the process of diffusive shock acceleration was investigated using an ensemble of independent test particles. In comparison to particle-in-cell simulations, the main advantage is that, aside from the model being easily adjustable, it is possible to correlate the acceleration process to the spatial diffusivity of the injected particles.

When the energy of the total particle ensemble is investigated, one finds power spectra being formed over time in all simulations. Independent of which model turbulence was applied, i.\,e. for both the magnetostatic and the \alfven-wave model of turbulence, a spectral index of close to $-2$ was found for parallel shock configurations. (For non-relativistic particles, the spectral index is accordingly reduced to $-1.5$.) This value equals the one predicted by the theory of first-order Fermi acceleration. In both simulations, particles remain within the acceleration region over long times and the total energy of the particle ensemble keeps growing. However, the power spectrum establishes itself quickly and only the high-energy tail changes toward the end of the simulation. The diffusivity of the particles over time indicates that the ballistic regime is not left within the maximum simulation time. This is due to the continued acceleration of some particles. To exclude this effect one either needs to apply even higher simulation times, or manually exclude particles from the acceleration region.

The simulation with a full model of \alfven waves, including their electric field components, illuminates the interplay of diffusive shock acceleration and stochastic acceleration. This has some minor influences on the resulting energy spectrum, leading to higher scattering of the data around the numerical fit corresponding to a spectral index of, in this case, $-1.9\pm 0.1$. Qualitatively however, no large differences are otherwise apparent.

The constructed magnetic fields of the two applied models of turbulence differ in the magnitude of the magnetic field downstream of the shock. For the magnetostatic model, the perpendicular components are scaled manually by the compression ratio, whereas the theory of \citet{schlickeiser2002cosmic} for \alfven wave transmission through parallel shocks leads to a more than twice as strong downstream  magnetic field. As such, with both models leading to comparable results for the power spectral indices, one \hl{might} argue that, despite the increased complexity of the \alfven-wave model of turbulence is unnecessary.
\hl{However, the general behavior is different with regard to the evolution of the coupled wave-particle system. In the presence of self-generated Alfv\'en waves, the relative contribution to the overall acceleration of the shock and the wave field can vary and lead to a wrong estimation of shock parameters if one were to neglect the influence of these waves. This effect may therefore complicate the interpretation of astrophysical gamma radiation in terms of shock acceleration \citep{aha04:snr}. This subject is currently under active investigation, and results will be presented in due time.}
In addition, note that the model used here is strongly restricted by the underlying theoretical understanding of \alfven-wave transmission, which is so far only valid for parallel shock configurations.

In addition, the work presented here investigated the shock acceleration process on a non-relativistic particle ensemble with a Maxwell-Boltzmann distributed rigidity injection profile. In conformance with the findings of \citet{caprioli2014simulations}, power spectra with spectral indices of close to $-1.5$ are obtained, and no injection problem is apparent. Similar to the relativistic simulations, which used a constant rigidity injection profile, the acceleration process becomes less efficient for oblique shocks. For a perpendicular shock then, a spectral index of ca. $-2$ is found. \citet{caprioli2014simulations}, too, reported a significant decrease of the acceleration efficiency for quasi-perpendicular shocks with angles above \ang{45}. Their results also show that parallel shocks are most efficient at accelerating particles, compatible to our findings. But, for quasi-parallel shocks, they also report a drop in acceleration efficiency, which cannot be seen in the results presented here, where no significant change in the spectral index is observable for oblique shocks below \ang{45}.

A limitation of the results presented here is the slab geometry that was used in the magnetostatic model of turbulence applied throughout the simulations presented in this work. This was chosen to ensure both the applicability of the \alfven wave transmission and reflection and the comparability with the magnetostatic model. However, a more realistic model might have a potentially large influence on the quasi-perpendicular shock simulations, because this geometry is symmetric in the directions perpendicular to the magnetic background field. Thus, in the limit of a perpendicular shock, the total turbulent magnetic field is static, even when it flows with the ambient plasma upstream or downstream of the shock. As such, particles only see the increase of the magnetic field strength, but are otherwise unaffected by the shock front. Future work certainly has to conduct the simulations presented here for different turbulence geometries, in order to investigate the effect this has on the acceleration process. Potentially, a more sophisticated geometry, such as the one proposed by \citet{goldreich1995toward}, will increase the diffusivity along the shock normal for quasi-perpendicular shocks and thereby improve the acceleration efficiency, increasing the number of high-energy particles and yielding higher spectral indices in the energy spectrum of the particles.

In conclusion, it has been shown that, even with the limited configurations used by the test-particle simulation presented in this work, reliable results can be obtained. This lays the necessary foundation for future studies. Most notably, the effect of the turbulence geometry on the acceleration efficiency of quasi-perpendicular shocks should be investigated, as mentioned earlier. Additionally, the influence of many other simulation parameters has to be checked. These parameters include, but are not limited to, the compression ratio, the shock speed, as well as the energy spectrum of the magnetic turbulence and the relative turbulence strength. Furthermore, if an extension of the \alfven wave transmission theory of \citet{schlickeiser2002cosmic} to oblique shocks can be derived, then it should be tested here as well. Also, a more sophisticated model of the background magnetic field, e.\,g., following the research by \citet{uya02:snr,eri11:acc}, could be included. Finally, one should lessen the simplifying assumptions on the shock front itself. Instead, a three-dimensional model of an expanding shock wave of finite thickness, potentially including precursor instabilities and/or Rayleigh-Taylor instabilities, could be used. In such a model, the shock front will have a varying obliqueness toward the magnetic background field, with so-far unexplored effects on the final upstream particle energy spectrum.

\begin{acknowledgements}
MW thanks Dieter Breitschwerdt and the ZAA for support. RCT acknowledges helpful discussions with Bram Achterberg and Fathallah Alouani Bibi.
\end{acknowledgements}



\end{document}